# A New Approach for the Interpretation of the Dynamic and Mechanical Properties of Polymer Nanocomposites Above and Below the Glass Transition

*Georgios Kritikos\* and Kostas Karatasos*

Laboratory of Physical Chemistry, Department of Chemical Engineering,

Aristotle University of Thessaloniki,

54124 Thessaloniki, Greece

## ABSTRACT

In this work, we present a new model for the interpretation of the local dynamic behavior and the mechanical reinforcement mechanism in polymer nanocomposites. The temperature dependence of the dynamics in the glassy region is described by a new equation which assumes an Arrhenius component in the cooperative diffusion. By doing so, a characteristic temperature which can be identified as the glass transition temperature ($T_g$) emerges, while an additional parameter for the extension of the super-Arrhenius region ($\delta_g$) is incorporated. Based on thermodynamic arguments, the dynamical heterogeneities are then related to structural heterogeneities in a manner consistent with the idea of a sigmoidal shape in the cohesion energy [*Kritikos G. Polymer (UK) 2014, 55 (18), 4658–4670*]. Incorporation of this temperature dependence of the cohesion energy to a Sanchez-Lacombe Equation of State, results in a sound description of the experimental temperature and pressure dependence of the density. Moreover, comparison with experimental data shows that the enhancement of mechanical properties in polymer nanocomposites can be associated with the extent of the glassy region.

*Keywords*: Glass transition; Polymer dynamics; Nanocomposites; VFT equation; Mechanical reinforcement.

**INTRODUCTION**

Polymer nanocomposites have been the subject of intense study for several decades now by both academia and industry [1–4]. Among the first applications of nanotechnology in polymers, were the nanocomposite elastomers [3–9]. In general, inclusion of a small percentage of nanofillers in the polymer matrix may result in enhanced mechanical and electrical properties compared to the pristine polymers [3–6]. In composites used for energy applications, the presence of the filler moieties can promote ionic diffusion while at the same time improve the mechanical stability of the materials [10,11]. Nevertheless, the prediction of the behavior of the nanocomposite polymers in a wide temperature range is limited by our understanding on fundamental issues in soft matter, such as glass transition, mechanical reinforcement mechanism and semi-crystallization.

To date, to our knowledge, no universal model exists, such as an equation of state (EoS), capable of describing the behavior of soft matter under conditions of cooperative diffusion in multiphase environments. Instead, interpretation of the observed properties is usually based on kinetic arguments [12–14]. Even the thermodynamic model by Gibbs – DiMarzio (GD) [15], treats the glass transition as a kinetic manifestation of an underlying thermodynamic glass transition, which takes place at a lower temperature where the conformational entropy becomes zero. Recent thermodynamic approaches, like the generalized entropy theory (GET) [16], that combine ideas from the Adam – Gibbs (AG) [17] theory and the lattice cluster theory (LCT) [18], support the existence of a cooperativity in the diffusion, in the super-Arrhenius (glassy) region. Based on the GET the cooperative rearranging regions (CRRs), introduced by the AG model, appear as string-like clusters.

In polymeric systems, a common experimental observation is that the activation energy in the super-Arrhenius region of glassy dynamics increases dramatically as the glass transition, $T_g$, temperature is approached from above. The temperature dependence of the diffusion coefficient near this region is commonly described, by the empirical Vogel-Fulcher-Tammann (VFT) equation [19–21], which, however, does not take into account the extent of the super-Arrhenius/glassy region. The relaxation process above $T_g$, identified as $α$-relaxation, follows a VFT-like behavior [22–24] attributed to the cooperative characteristics of the diffusion mechanism. Below $T_g$, the glassy material follows a relaxation procedure towards a uniform density, referred to as aging [25,26]. Within the glassy region, experimental techniques such as dielectric relaxation spectroscopy, continue to detect mobility [27]. In this region the principal motional mechanism, $β$-relaxation, is described by an Arrhenius behavior of constant activation energy. In multicomponent systems, such as polymer nanocomposite materials, experimental works [3,6,28–30] probing dynamics of the bound layer [31–34] or examining confined polymer films, indicate the presence of an Arrhenius type of dynamics even above, but still close to $T_g$.

Under cyclic loading, mechanical properties of polymer nanocomposites deteriorate according to the "Payne" and "Mullins" effects [35,36]. The explanation of the observed behavior is based on two models. One model [29,37] considers that a solid-like nanofiller phase is responsible for the increased elastic modulus of the nanocomposites. According to this approach, the observed behavior is attributed to the formation of a network of filler particles, immersed within the polymer matrix. Application of shear stress can destroy this network reducing thus the modulus values. The second model [29,38] is based on the assumption that glassy bridges are formed between neighboring filler particles. This approach is consistent with arguments supporting the existence of a distribution of $T_g$'s in thin polymer films [1,39,40]. Deformation may disrupt these glassy polymer bridges leading to a lower degree of mechanical stability. In such processes, where the stress transfer mechanism plays a central role, relevant studies on polymer/graphene nanocomposites have highlighted the significance of the nanosized dimensions of the filler particles [10,41].

A few years ago, a new idea was presented [32] based on thermodynamic arguments for the interpretation of the influence of the nanofillers in the mechanical properties in

poly(dimethyloxane) PDMS/Silica systems. In contrast to previous studies on similar systems, where the picture of more than one glass transition temperatures was suggested [30], an alternative view was proposed based on the presence of a single $T_g$ temperature, accompanied by a stronger transition of the bound polymer layer to a rigid amorphous phase [32,42]. This approach could account for the fact that although Dielectric Relaxation Spectroscopy (DRS) and Differential Scanning Calorimetry (DSC) experiments in nanocomposites detected a reduced dielectric strength and a heat capacity step respectively, compared to the pristine polymer, they did not detect an additional dynamic or calorimetric glass transition associated with the nanofiller [6,29,30,33,40,42]. Based on this "bound-layer model" [32,42,43], a decoupling between the adsorbed and the rest of the polymer could be described [31,33,42,44]. According to this picture, even in the case of highly attractive nanoparticles, the presence of an immobilized polymer layer may result in neutral interactions between the polymer and the nanofiller (i.e., essentially setting the Flory-Huggins parameter [45,46], $\chi$, close to zero [42]).

The aforementioned mean-field model [10,32,42] treats the glass transition as an entropy-driven phenomenon. At sufficiently high temperatures, the diffusion is considered to have non-cooperative characteristics and is described by a constant activation energy. As the temperature drops, the free volume reduces and the molecular diffusion needs to adopt cooperative characteristics in order to avoid the entrapment into a cage [47] which would result in a large reduction in the entropy [32,48]. In the case of extreme confinement, this entropic force becomes responsible for the decoupling between layers [3,6,42,49,50].

In the next section we present the details and the formalism of the proposed model aiming at the prediction of dynamical and mechanical properties of polymer nanocomposites. To this end, a VFT-like equation and an Equation of State (EoS), which incorporate a parameter for the extension of the glassy region, are described. In the sections to follow a comparison of the model with pertinent experimental data is also presented. The paper concludes with the main findings and the prospective impact of the work presented, toward a better understanding of the dynamic and mechanical behavior in polymer nanocomposites.

**MODEL DESCRIPTION**

**Cohesion Energy in the Glass Transition Region.**

Our approach is based on the assumption that an Arrhenius component contributes in the diffusion, even in the super-Arrhenius region. Namely, we propose that the diffusion retains an Arrhenius component of constant activation energy even when entering the region of glassy dynamics. This idea is consistent with a picture according to which the cooperative molecular diffusion creates areas of excess free volume where free diffusion is allowed, while the cost for the existence of such areas results in an increase in the activation energy. Under these conditions local density relaxation can be realized via a composite Debye process described, e.g., by a modified KWW equation [24,51], which consists of one simple and one stretched exponential.

To better visualize this concept, we consider for simplicity a nanocomposite polymer system comprised by a homopolymer and monodisperse nanoparticles, where the latter are too large to diffuse and thus do not participate in the entropy of the system. Analogous cases of mobile nanoparticles or different kind of polymers can be treated according to the Sanchez – Lacombe theory [52] using combination rules that have been described in previous works [52–55]. We take [42] that the temperature dependence of the activation energy in the glassy region, $A^*$, follows the temperature dependence of the cohesion energy ($\varepsilon$) and can be described by an equation of the form:

$$A^* = A\left(g(\varepsilon_{ref}/\varepsilon) + 1\right), \quad (1)$$

where $A$ is the activation energy in the Arrhenius region at sufficiently high temperatures[56], $g$ is a function representing the non-Arrhenius part and $\varepsilon_{ref}$ is a reference value introduced for dimensionality reasons. The introduction of $g$ is required in order to reproduce the experimentally tested VFT behavior [22–24]. The cooperative diffusion can then be described as a bimodal process (i.e., consisting of an Arrhenius and a non-Arrhenius part). This cooperativity is related to structural heterogeneities [16,17,57], which are reflected in the cohesion energy of the entire system.

For the description of the temperature dependence of the cohesion energy, a previous idea of a varying mean-field [42,58] first presented in an EoS based numerical Self Consistent Field (EoS-nSCF) study [42] on polymer nanocomposites, will be adopted. Let us assume a lattice having a coordination number $z$. A coordination number of e.g., $z = 6$ would correspond to the cubic lattice, while as z increases the continuous (i.e, real) space, is approached. Let $V$ represent the entire volume of the system and $v$ a single lattice site. Then $N$ chains having $r$ segments each, occupy $Nr$ lattice sites [45,46]. In order to be able to describe the temperature dependence of the density, we also introduce a number of vacant sites ($N_{vac}$) [42,52,59,60]. Based on the above notation, the system's volume is given by $V = v(Nr + N_{vac})$, while the volume fraction of the polymer is described as, $\Phi = \dfrac{rNv}{V}$.

According to the traditional mean field approach [45,46], the nearest neighbor segments (of type $A$) in the liquid state interact with a constant cohesion/pairwise energy, $w_{AA}$. Then, the nonbonded energy on a lattice can be described as the product $rN\Phi\varepsilon_2$, where $\varepsilon_2 = zw_{AA}/2$. Following our past work [42], we substitute gradually (continually and randomly) part ($f$) of the $w_{AA}$ interactions with the lower energy level (solid state), $s_{AA}$, interactions. We assume that the entire glassy material occupies this energy level of $s_{AA}$, at the Vogel temperature $T_o$. Also, we define the energy parameter, $\varepsilon_1 = zs_{AA}/2$. In this way, the mean field energy in the glassy region (near $T_g$) is now given as [42]:

$$E = rN\frac{rNv}{V}\varepsilon, \tag{2a}$$

where $\varepsilon = \varepsilon_2 - f\varepsilon_2 + f\varepsilon_1 = \varepsilon_2 + f(\varepsilon_1 - \varepsilon_2)$. If we introduce the positive quantity $T^* = -\varepsilon/k$, that has units of temperature ($k$ is the Boltzmann constant), then the enthalpic factor can be described as:

$$E = -krN\frac{rNv}{V}[T_2^* + f(T_1^* - T_2^*)] \tag{2b}$$

As the cooperative diffusion takes place, a fraction of mobile regions, $f_2^*$, should retain the cohesion energy of high temperatures ($T_2^*$), while regions of the lowest energy level, $T_1^*$, will also start growing, attaining a total fraction of $f_1^*$. Therefore, based on the modified [10,42,58] mean field approach:

$$T^* = f_2^* T_2^* + f_1^* T_1^*, \tag{3}$$

where the fraction of the mobile regions is described as [10] $f_2^* = (T_1^* - T^*)/(T_1^* - T_2^*)$ and the fraction of the randomly distributed immobilized regions is given by $f_1^* = (T^* - T_2^*)/(T_1^* - T_2^*)$.

In the traditional VFT equation [19–21], the activation energy is considered inversely proportional to the temperature difference from the Vogel temperature, i.e. $1/(T - T_o)$. In our model we consider [10] $A^*$ to be inversely proportional to the fraction of the mobile regions, i.e:

$$A^* = A \frac{T_1^* - T_2^*}{T_1^* - T^*}. \tag{4}$$

In other words, in consistency with the VFT approach, at each temperature, $A^*$ is inversely proportional to the energy difference (temperature units) from the solid state (at Vogel temperature) normalized with the energy step $T_1^* - T_2^*$. Combination of eqs. 1, 3 and 4 renders the function $g$ as:

$$g(\frac{T_{ref}^*}{T^*}) = \frac{f_1^*}{f_2^*}, \tag{5}$$

where $T_{ref}^*$ is defined as $T_1^* - T_2^*$, while for the fractions $f_1^*, f_2^*$ it stands: $0 \leq f_1^* \leq 1$, $0 \leq f_2^* \leq 1$ and $f_1^* + f_2^* = 1$. This expression for the non-Arrhenius part attributes universal characteristics for every glassy material that enters the super-Arrhenius region. For the limiting behavior of eq. 5, at high temperatures i.e., at the Arrhenius region, $g \to 0$, at the solid state $g \to \infty$, while there is also a characteristic temperature at which $f_1^* = f_2^*$ and $g=1$.

For temperatures below this characteristic temperature, the non-Arrhenius component for the diffusion becomes inaccessible and the cooperative diffusion ceases. A collapse of the heat capacity of the system is then expected, which is compatible with the experimental observations of a heat capacity step and a change in the slope of the specific volume [10,42]. Therefore, this characteristic temperature at all the measurable aspects can be identified as the glass transition temperature, $T_g$.

Apart from the limiting cases, for other values of $f_1^*$ or $f_2^*$, function $g$ in eq. 5 can be approximated by an exponential growth with decreasing temperature (see Figure S1 of the supplementary material (SM) section). In order to quantify the behavior of $g$ in the temperature representation, an additional parameter, $\delta_g$, with units of temperature is introduced, as shown in eq. 6:

$$g(T) = \exp\left(\frac{T_g - T}{\delta_g}\right). \tag{6}$$

Parameter $\delta_g$ essentially relates to the extent of the heterogeneities (associated with the glassy behavior) above and below $T_g$, and is considered to be characteristic for each glassy system and the cooling rate used. The resulted temperature dependence of the activation energy will be checked against experimental data in a section to follow.

The new equation for the activation energy (i.e, the combination of eq 1 and eq 6) which incorporates the above ideas of cooperative diffusion can be introduced for the calculation of the relaxation time $\tau$ according to [10]:

$$\tau = \tau_o \exp\left(\frac{A}{T}\left[\exp\left(\frac{T_g - T}{\delta_g}\right) + 1\right]\right). \tag{7}$$

We will refer to eq. 7 as the KK [10] equation.

From eqs 1,4,6 it follows that: $\frac{T_1^* - T_2^*}{T_1^* - T^*} = \exp\left(\frac{T_g - T}{\delta_g}\right) + 1$. Solving with respect to $T^*$, yields the temperature dependence of the cohesion energy at the super Arrhenius region:

$$T^* = T_2^* + \frac{1}{1 + \exp\left(\frac{T - T_g}{\delta_g}\right)}\left(T_1^* - T_2^*\right), \tag{8}$$

where the factor $f = 1/\left(1 + \exp\left(\frac{T - T_g}{\delta_g}\right)\right)$ represents a sigmoidal switching function of temperature. We will refer to eq. 8, as the Solid to Liquid Glass (SLG) transition function, which in consistency to previous work [42] represents a two state switching function that allows a mean field description of the glassy region.

**Equation of State**

A thermodynamic approach to the glass transition should be describable by a relevant EoS. Based on the Flory-Huggins theory [45,46], entropy depends on the natural logarithm of the volume fraction (probability) of each entity as:

$$S = kN \ln\left(\frac{V}{rNv}\right) + k\frac{V-rNv}{v}\ln\left(\frac{V}{V-rNv}\right) \tag{9}$$

After incorporating SLG mean field ideas [42] regarding the enthalpic factor (eq. 2b), the Helmholtz free energy ($F = E - TS$) is given as:

$$F = -krN\frac{rNv}{V}[T_2^* + f(T_1^* - T_2^*)] - TkN\ln\left(\frac{V}{rNv}\right) - Tk\left(\frac{V-rNv}{v}\right)\ln\left(\frac{V}{V-rNv}\right) \tag{10}$$

We assume [10,42] that the replacement of the liquid interactions by the solid interactions on the lattice is made in such a way that the system retains the translational entropy as described by the Flory-Huggins theory [45,46].

Moreover, the pressure is defined as:

$$P = -\left(\frac{\partial F}{\partial V}\right)_{N,T}$$

Based on the above statistical mechanical representation, we may express the EoS at the glassy region as:

$$\frac{Pv}{kT} + \left(1 - \frac{1}{r}\right)\left(\frac{\rho}{\rho_o}\right) + \ln\left(1 - \frac{\rho}{\rho_o}\right) + \left[\frac{T_2^*}{T} + \frac{T_1^* - T_2^*}{\left(1 + exp\left(\frac{T-T_g}{\delta_g}\right)\right)T}\right]\left(\frac{\rho}{\rho_o}\right)^2 = 0 \quad, \tag{11}$$

where $\rho_o$ is the density in the Vogel temperature, $\rho$ is the density at each temperature and $\frac{\rho}{\rho_o} = \frac{rNv}{V}$. Assuming that $\rho$ is calculated in $g/cm^3$, the molecular weight (MW) in $g/mol$ and $v$ in Å$^3$, then $r$ is equal to $MW / (0.6022\rho_o v)$ [52]. We will call the new EoS as SLG-EoS. It encapsulates a coupling between dynamics and thermodynamics in the glassy region.

The introduction of areas with different cohesion energy and pronounced pressure fluctuations ($kT^*/v$) is compatible with the idea of the CRRs [17]. We consider them to be diffusible when diffusion is allowed. For this reason, they should contribute to the entropy of the system. Since the SLG switching function (eq. 8) is a free energy term, then the probability to find a polymer segment in an immobilized region compared to the case where it is located in a mobile region (of $T_2^*$ interaction energy) is, $\exp[-\dfrac{rN\dfrac{rNv}{V}f(T_2^* - T_1^*)}{T}]$. It follows that the CRRs contribute to the entropy by a factor of: $k\dfrac{rN\dfrac{rNv}{V}f(T_1^* - T_2^*)}{T}$. This implies that the entropy increases when the number of contacts on the lattice, between mobile and immobile segments increases. It is, therefore, expected that the smallest possible immobilized islands ("droplets of solid state") are formed. This description, is consistent with results concerning the temperature dependence of the pair distribution function [49] and with findings presented in a relevant review paper [57]. We point out that derivation of the SLG [42] two state function (eq. 8), does not require the formation of string-like CRRs [16].

**RESULTS AND DISCUSSION**

**Testing the Model against Experiment.**

In the sections to follow we hereby attempt to test the proposed model against available experimental results referring to diverse systems, in order to demonstrate its ability to capture generic characteristics of the dynamics related to glass transition phenomena in polymer composites. Although the list of the experimental systems that were checked was far from being exhaustive compared to the existing literature in the field, we believe that the cases studied can be representative for a far larger number of systems bearing similar characteristics.

In Figure 1, we fit the KK equation (eq. 7), with experimental results on Polystyrene (PS) nanocomposites, as published by Schönhals et al. [27]. The nanocomposites were prepared [27] by the inclusion of Phenethyl-POSS into the PS matrix by a solution blending technique. In this PS system the insertion of the nanofillers, resulted in a reduction of the calorimetric $T_g$. The experimental values as evaluated by the DSC technique were: 373.1, 362.3, 358.0, 351.7, 343.0 and 325.8 K for the PS000, PS005, PS010, PS022, PS032 and PS038, respectively [27]. POSS is considered [27,62,63] as the smallest possible silica particle and thus can act as a plasticizer for the polymer diffusion, by increasing the free volume of the nanocomposite polymeric system compared to the net polymer.

We remind that according to our model, $T_g$ is considered as the temperature where the structural heterogeneities formed due to the lack of free volume, lead to a balance between the fractions of the mobile and immobile regions, which essentially translates to a normal distribution in the molecular volumes [10]. The increase of polymer MW causes a $T_g$ increase as the fraction of free ends decreases and so does the total free volume (i.e, density increases) [64]. In case of the presence of high MW nanoparticles, which diffuse in the polymer matrix, a similar increase in the $T_g$ can be observed [65,66]. On the other hand, in case of nanoparticles that cause a plasticization effect, the glassy region described by both $T_g$ and $\delta_g$, should be shifted to lower temperatures. Moreover, if the nanoparticles do not diffuse, even in case of polymer adsorption a reduction/stabilization of the $T_g$ and an increase of $\delta_g$ can be observed [6,29,42].

The solid lines in Figure 1, show the result of fitting the experimental data with the KK equation. Frequencies ($F$) correspond to $1/(2\pi\tau)$, where $\tau$ is the relaxation time appearing in eq. 7. At all percentages in POSS, the KK equation fits properly the temperature dependence of the dynamics, depicting a behavior that is similar to the traditional VFT equation. In addition, as shown in Table 1, the experimentally [27] observed reduction in the $T_g$ can be reproduced, while an estimation regarding the region $\delta_g$, where the structural heterogeneities are present, can be made. The error bars reflect the fact that somewhat different sets of parameters may also provide an acceptable fitting. Incorporation of statistical weights based on the error margins of the experimental values, could provide a more accurate description of the experiment.

Concerning the activation energy, $A$, the value for bulk PS is $-2905 \pm 400$ K. After fitting the $\beta$ relaxation as presented in Figure 5 of Ref [27] (not shown here) to an Arrhenius equation of the form $\tau = \tau_o \exp\left(\dfrac{A}{T}\right)$, we have estimated an $A$ of $-3400 \pm 200$ K. Taking into account the uncertainty in the evaluation of *A,* we may conclude that both KK and Arrhenius equations agree in the estimation of the activation energy. Since below $T_g$ there is a consensus about the Arrhenius characteristics of the *β* relaxation [56], this agreement is considered as an additional check of our model, which assumes a single parameter for the activation energy for the entire temperature range (see eq. 1).

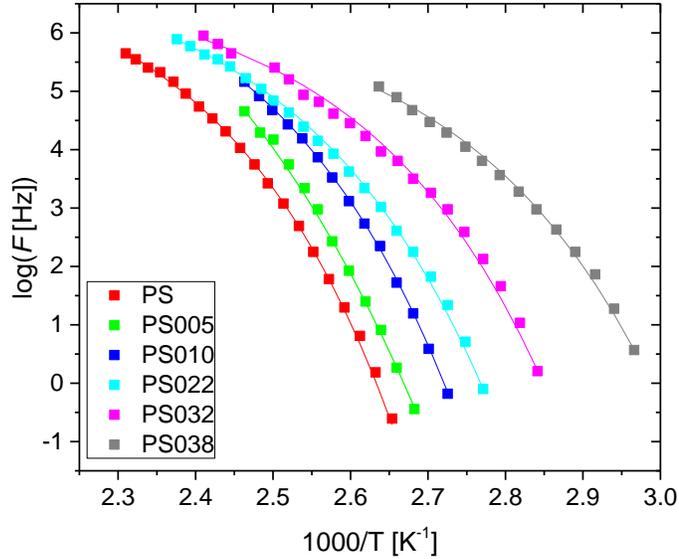

Figure 1. Relaxation frequency *F* vs inverse temperature of the α mechanism, for various percentages of Phenethyl-POSS in the PS matrix, as presented in Ref. [27]. The solid lines represent fits according to the KK equation.

|  | $T_g$ [K] | $δ_g$ [K] | A [K] | $F_o$ [Hz] |
|---|---|---|---|---|
| **PS** | 371±5 | 28±5 | -2905±400 | (3.7±10) x$10^5$ |
| **PS005** | 367±5 | 41±10 | -3185±300 | (5.6±10) x$10^6$ |
| **PS010** | 355±5 | 27±5 | -3225±300 | (1.6±10) x$10^6$ |
| **PS022** | 353±5 | 27±5 | -2626±300 | (3.1±10) x$10^5$ |
| **PS032** | 349±5 | 26±5 | -2040±300 | (7.2±10) x$10^4$ |
| **PS038** | 325±5 | 25±5 | -2215±300 | (1.5±10) x$10^4$ |

Table 1. Values of the fitting parameters of eq. 7 to the experimental results presented in Figure 1. $F_o$ is equal to $1/(2\pi\tau_o)$.

The thermodynamic description is based on the concept that the activation energy of the dynamic processes is related to the cohesion energy. This can be corroborated if, for the same polymer (i.e. PS), the temperature dependence of the density can be also described by the proposed model. In Figure 2, we compare the temperature dependence of the density, as described by the SLG EoS (eq. 11), to experimental data in a wide temperature range, both above and below $T_g$. The experimental results [67] correspond to PS nanocomposites with 2 wt% in clay. In such PS composites an improvement in mechanical properties has been observed [67,68], accompanied by a stabilization [27] or an increase of $T_g$ [68]. In addition, comparison is made with the experimental results covering a wide pressure range from 0.1 MPa to 130 MPa. The fitting parameters are shown in Table 2.

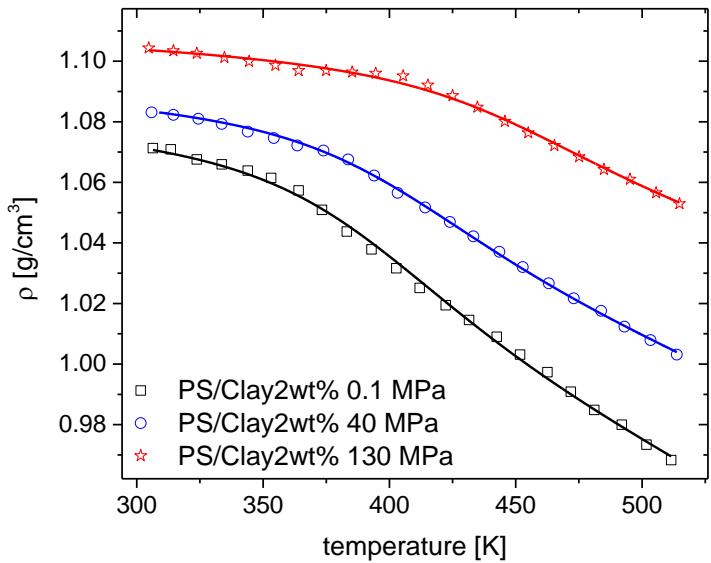

Figure 2. Temperature dependence of the density for PS/clay 2 wt% nanocomposites. The black squares, blue circles and red stars denote the experimental measurements [67], while the solid black, blue and red lines represent the fitting with the SLG EoS (eq. 11), for pressures of 0.1, 40 and 130 MPa, respectively.

| PS/Clay 2wt% | $T_g$ [K] | $\delta_g$ [K] | $T_1^*$ [K] | $T_2^*$ [K] | $\rho_0$ [g/cm$^3$] | $v$ [Å$^3$] | MW [g/mol] |
|---|---|---|---|---|---|---|---|
| 0.1 MPa | 380±2 | 30±5 | 1016±50 | 650±20 | 1.080±0.01 | 25788±500 | (1±100)x10$^5$ |
| 40 MPa | 393±2 | 30±5 | 1016±50 | 650±20 | 1.090±0.01 | 95±10 | (1±100)x10$^5$ |
| 130 MPa | 440±2 | 30±5 | 1016±50 | 650±20 | 1.107±0.01 | 48±10 | (1±100)x10$^5$ |

Table 2. Values of the fitting parameters of eq. 11 to the experimental results presented in Figure 2.

The proposed EoS provides a very good description of the experimental results allowing for an accurate description of the compressibility and the thermal expansion coefficient. Note that in Table 2 the parameters $\delta_g$, $T_1^*$ and $T_2^*$, which are material-specific, were kept constant. Also, we quote that the MW corresponds to an average molecular weight which describes both, the polymer chain and the nanofiller.

An additional test of the SLG-EoS is shown in Figure S2 of SM, where the experimental density temperature dependence [69,70] of poly(acrylic acid) (PAA) is fitted. The description is very good allowing for an extrapolation at lower and at higher temperatures. The estimated $T_g$ was 400 ± 5 K, a value which is close to the one quoted for the simulated PAA (412 ± 5 K) [10] and the experimentally measured value of 401 K [69,70]. The results support the analysis that was presented [10], on the framework of the proposed model.

The performance of the proposed SLG-EoS [42] in describing such thermodynamic data, attests to the significance of the SLG two-state function [10,42]. A recent work [71] on Coarse Grained (CG) simulations which considered such sigmoidal switching function in the derivations of the non-bonded interactions, supports this idea, as well. By defining the cohesion energy as a function of $T_g$, efficient transferable Coarse Grained (CG) potentials at lower temperatures and higher MWs can be obtained, establishing a stricter connection with the structure and dynamics of the corresponding atomistic simulations.

**Bound Layer Dynamics.**

Experimental studies [3,6,28–30,42] on the dynamics of a category of nanocomposites, identified above $T_g$ the existence of an additional to the $\alpha$-relaxation mechanism, which exhibited an Arrhenius behavior and was attributed to the bound/adsorbed/restricted layer. In previous computational studies [42,72] we have followed the transition of the bound layer to a rigid amorphous phase ("dead layer") in poly(ethylene) (PE), which is considered a model polymer for simulations with a simple, reliable force field, allowing efficient Monte Carlo (MC) sampling. The solidification was located almost 100 K above $T_g$ and was found to be significantly stronger compared to the main $T_g$ transition. This analysis provided an interpretation for the observed Arrhenius dependence of the adsorbed layer dynamics, on the basis of an extension of the super-Arrhenius region.

In semi-crystallized polymer nanocomposites, it has been observed [30,42,73] that after annealing above $T_g$, the heat capacity step or the dielectric strength of the total mobile fraction that undergoes glass transition, shows a tendency for stabilization, independently of the degree of crystallization or of the nanofiller content. Therefore, from a thermodynamic perspective, semi-crystallization can be treated [10,42,49] as an inherent tendency of glassy materials, such as polymers, to extend their super-Arrhenius region in order to avoid the Kauzmann paradox [74] of a steep entropy drop. In crystallizable polymers, due to hysteresis phenomena, a discontinuity in the reduction of the free volume between the $T_m$ and the crystallization temperature ($T_c$), is observed [42].

In a dielectric spectroscopy study (Figure 3) on poly(dimethylsiloxane) (PDMS) nanocomposites, presented by Kremer et al. [3], it is clear that in the pristine polymer the so-called $\alpha_c$ mechanism, which refers to the mobile polymer between the crystalline nanoregions, is broader than the conventional $\alpha$-relaxation process. In that study the $\alpha$-relaxation mechanism was monitored after heating a quenched polymer sample, bypassing thus the crystallization. A spontaneous extension of the super-Arrhenius region was recorded [3,42].

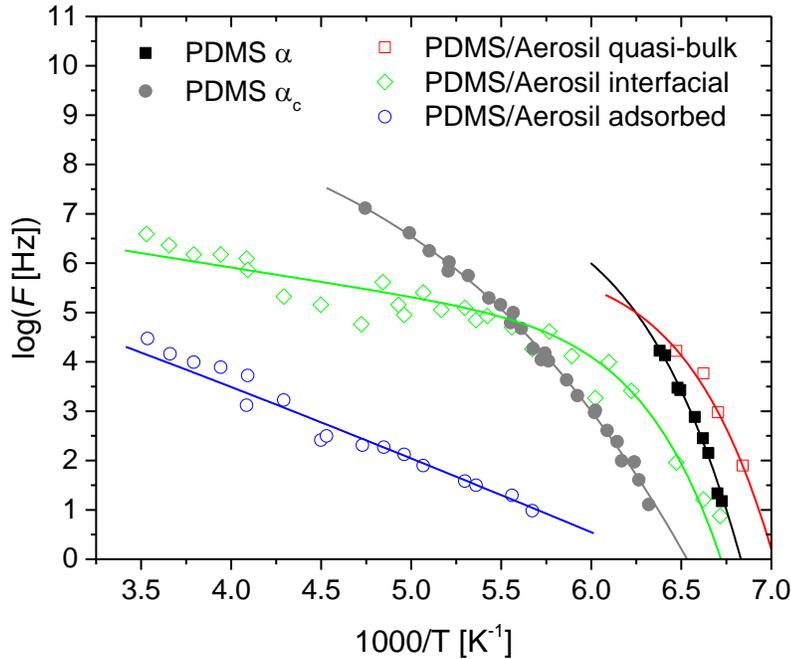

Figure 3. Relaxation frequency $F$ vs inverse temperature of $\alpha$ and $\alpha_c$ relaxation mechanisms of pure PDMS, of quasi-bulk, of interfacial and of adsorbed layers, in PDMS/AeroSil 0.5 vol% nanocomposite systems, as presented in Ref. [3].

Figure 3, portrays results for both, the pure polymer and the PDMS/hydrophilic AeroSil composites. In all systems, the KK equation provides a good description of the temperature dependence of local dynamics. Table 3, lists the corresponding fitting parameters. The predictions regarding the $T_g$ temperature are close (slightly higher) to the one presented in the experimental study [3]. An almost 6% difference is reasonable since the estimation of the $T_g$ values in the experimental work, was based on a fitting to the Williams-Landel-Ferry (WLF) equation [75]. Interestingly, as shown in Table 3, both, the $\alpha_c$ and the adsorbed layer relaxation mechanisms assume increased values of the $\delta_g$ parameter.

|  | $T_g$ [K] | $\delta_g$ [K] | A [K] | $F_o$ [Hz] |
|---|---|---|---|---|
| **PDMS α** | 160±10 | 12±5 | -335±50 | (9.7±10) x$10^3$ |
| **PDMS α$_c$** | 160±10 | 40±5 | -800±50 | (1.2±10) x10$^5$ |
| **quasi-bulk** | 160±10 | 10±5 | -144±50 | (8.8±10) x10$^2$ |
| **interfacial** | 150±10 | 10±5 | -583±50 | (3.8±10) x10$^3$ |
| **adsorbed** | 140±10 | 300±50 | -624±50 | (2.2±10) x10$^3$ |

Table 3. Values of the fitting parameters of eq. 7 to the experimental results presented in Figure 3. $F_o$ is equal to $1/(2\pi\tau_o)$.

Another test case that was examined, refers to the behavior of Nitrile butadiene rubber (NBR)/silica nanocomposites, as described by Xu et al [6]. Rubber-based nanocomposites are among the most well studied composite systems due to their numerous applications [4,9,76–80]. Inclusion of a small fraction of silica particles in butadiene rubber may result in dramatic mechanical reinforcement, i.e. around 2 orders of magnitude increase in the shear modulus, $G$ [38]. In Figure 4, we examine the performance of the proposed model in describing the behavior of the systems studied in Ref [6]. Both the mobile and the restricted polymer layers, as identified in the relevant experiment, have been fitted with the KK expression. The corresponding fitting parameters are listed in Table 4.

The results of the fit are very good for both, the restricted (res) and the mobile polymers layers. The glass transition experimental values were [6] 238, 237, 235 and 238 K for the pure NBR, the A200(0.26), the A380(0.30) and the R974(0.30) respectively. The $T_g$ values evaluated by the KK equation (Table 4) are in close agreement with those experimentally measured [6]. In addition, we have tried to estimate the $T_g$ values based on the temperature dependence of the restricted/bound layer. Comparison of the $T_g$s for the restricted fraction with the ones describing the mobile layer, shows that eq. 7, within the error margin, predicts practically indistinguishable glass transition temperatures, although the uncertainty in this calculation is larger compared to the one based on the mobile layer. The overall analysis verifies the tendency for stabilization of $T_g$, in accordance to experimental results [6]. A marked difference, however, is observed between the estimated $\delta_g$ values describing the mobile and the restricted fractions. A similar trend that was observed previously in the PDMS/silica samples (see Table 3) is noticed in these nanocomposites, as well. The reinforcement of the nanocomposite elastomers, which experimentally was found to grow with the filler content, appears to be correlated with an extension of the super-Arrhenius region, as described by the $\delta_g$ parameter.

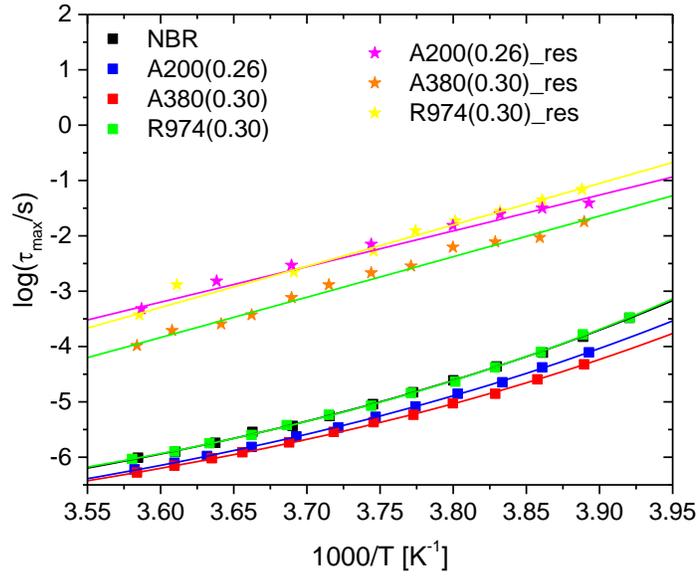

Figure 4. Maximum relaxation times $\tau_{max}$ of the restricted (res) and the mobile fractions as a function of the inverse temperature, for neat and nanocomposite (compound) NBR, as presented in Ref. [6]. The solid lines through the points indicate the fittings according to eq. 7.

|  | $T_g$ [K] | $\delta_g$ [K] | A [K] | $\tau_o$ [s] |
|---|---|---|---|---|
| **NBR** | 240±5 | 24±5 | 1469±100 | (4.35±10) x$10^{-6}$ |
| **A200(0.26)** | 240±5 | 25±5 | 1382±100 | (5.00±10) x$10^{-6}$ |
| **A380(0.30)** | 230±5 | 22±5 | 1851±100 | (1.22±10) x$10^{-6}$ |
| **R974(0.30)** | 250±10 | 24±5 | 1041±100 | (1.92±10) x$10^{-5}$ |
| **A200(0.26)_res** | 220±20 | 400±100 | 2600±500 | (1.07±10) x$10^{-9}$ |
| **A380(0.30)_res** | 220±20 | 400±100 | 2947±500 | (5.45±10) x$10^{-11}$ |
| **R974(0.30)_res** | 220±20 | 400±100 | 3012±500 | (6.08±10) x$10^{-11}$ |

Table 4. Values of the fitting parameters of eq. 7 to the experimental results presented in Figure 4.

To further corroborate the hypothesis of a coupling between the degree of mechanical reinforcement and the extension of the glassy region in polymer nanocomposites, we have compared the predictions of the proposed model to additional relevant experimental data [29]. These data refer to nanocomposites of poly(2-vinylpyridine)(P2VP) with silica nanospheres which were characterized using various experimental and computational techniques [29]. A reinforcing action of the silica nanoparticles was found, while a transition to an Arrhenius-like behavior was observed at high silica loadings close to 31 vol%. Figure 5 shows the result of the fit with the proposed model, to the shifting factors, $a_T$, used to construct the viscoelastic master curves [29]. The parameters which describe the solid lines of the KK expression, are shown in Table 5.

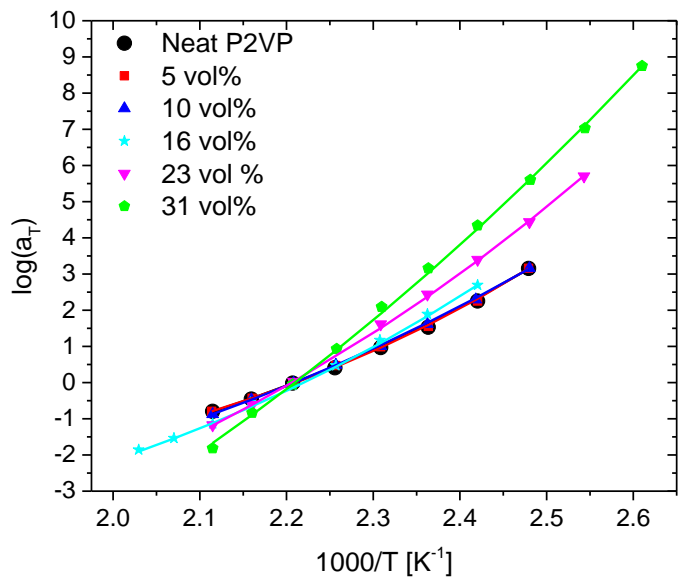

Figure 5. Shift factors, $a_T$, used for the construction of the experimental viscoelastic master curves, as a function of inverse temperature for pristine P2VP and for the examined nanocomposites, as presented in Ref. [29].

|  | $T_g$ [K] | $\delta_g$ [K] | A [K] | $\tau_o$ [s] |
|---|---|---|---|---|
| **P2VP neat** | 360±10 | 69±10 | 3094±100 | (1.8±10) x$10^{-4}$ |
| **P2VP 5vol%** | 360±10 | 61±10 | 3152±100 | (2.0±10) x$10^{-4}$ |
| **P2VP 10vol%** | 360±10 | 91±10 | 3103±100 | (8.8±10) x$10^{-5}$ |
| **P2VP 16vol%** | 369±10 | 80±10 | 3386±100 | (3.5±10) x$10^{-5}$ |
| **P2VP 23vol%** | 395±20 | 99±10 | 3224±100 | (1.3±10) x$10^{-5}$ |
| **P2VP 31vol%** | 360±10 | 122±5 | 5733±1000 | (8.4±10) x$10^{-5}$ |

Table 5. Values of the fitting parameters of eq. 7 to the experimental results presented in Figure 5.

In this experimental work, a stabilization of the $T_g$ around 375 K (inset of Figure 4 of Ref. [29]) for various filler contents, was recorded. The results for the estimated $T_g$ values listed in Table 5 are consistent with this tendency if the error of the evaluation is taken into account. As it was noticed in the systems examined earlier, the increased value of the parameter $\delta_g$ at high volume fractions of the silica nanoparticles, is commensurate with the experimentally observed increase in the modulus at high silica content [29]. The extension of the glassy region in these systems is also indicated by the lower Vogel temperature and the stabilized $T_g$, as the filler content increases, according to the authors' analysis [29]. Although the results for the shifting factor do not allow for a layer analysis as shown previously [3,6], it appears plausible to attribute the extension of the super-Arrhenius region to the bound layer transition. In extreme confinement conditions a static cooperativity is essentially present, which can be manifested by a stronger transition of the adsorbed layer, while the middle layer exhibits faster dynamics and more fragile glass transition characteristics [3,6,40,42]. The overall glassy region extends in order to include the stronger transition of the bound layer.

**Origin of Mechanical Reinforcement.**

In previous works [10,49] we have addressed the issue concerning the origin of the mechanical stress responsible for the enhanced mechanical properties in nanocomposite/confined glassy materials. In a molecular dynamics (MD) simulation study [10] on poly(acrylic acid)/graphene (PAA) nanocomposites we have examined in a layer-resolved analysis, the possible origin of the increased shear modulus, which in case of polyelectrolyte membranes could also induce increased ions' transport properties.

Detailed analysis through MD simulations in stress – strain experiments [10,49] did not reveal the existence of "glassy bridges" [38] between the nanofillers, even in cases of high filler contents, or short interparticle distances. This was also confirmed by EoS-nSCF [32] and MC investigations [72] concerning the order parameter of the intermediate polymer layer between opposite plates. The computational and experimental studies on various nanocomposite polymers [10,32,40,49,72], supported the idea that the origin of the mechanical reinforcement mechanism is located in the middle layer, which exhibits liquid mobility. In some cases of extreme confinement, an Arrhenius component can be isolated, while the heterogeneity is expressed with intense pressure fluctuations that resist to further deformation [10,49].

In filled elastomers, the relation between the viscosity ($\eta$) and the shear modulus ($G$) can be described by an equation of the form $\eta^1/\eta^0 = G^1/G^0$, (see chapter 7, in Ref. [81]), where index 0 stands for the bulk material and index 1 stands for the filled one. Assuming that the viscosity follows the KK temperature dependence, then the ratio $G^1/G^0$ can be described as:

$$\frac{G^1}{G^0} = \frac{\exp\left(\frac{A^1}{T}\left[\exp\left(\frac{T_g^1 - T}{\delta_g^1}\right) + 1\right]\right)}{\exp\left(\frac{A^0}{T}\left[\exp\left(\frac{T_g^0 - T}{\delta_g^0}\right) + 1\right]\right)} \quad (12)$$

where $A^i$, $T_g^i$ and $\delta_g^i$ ($i=0,1$) are the activation energies, the glass transition temperatures and the temperature extension parameters (for the bulk and the nanocomposite) respectively. We mention that a similar temperature dependence of the shear modulus has been also proposed by others [82]. Based on the analysis of experimental results presented above, we may assume that in cases of nanocomposites where the filler does not diffuse, the activation energy and glass transition temperature can be constant, independent of the filler content.

In Figure 6 we plot the ratio, $R= G^I/ G^0$, of the NBR/silica nanocomposites [6] for different $\delta_g^I$ values. We assume, $T_g^i =240$ K, $A^i=1400$ K and $\delta_g^0=25$ K.

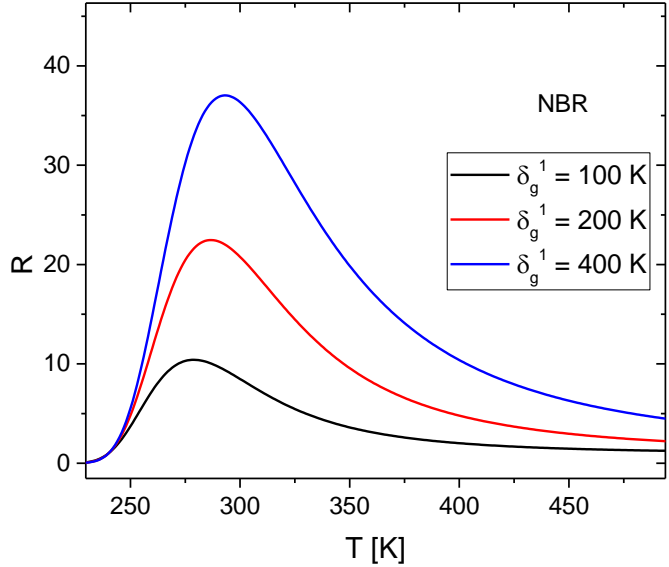

Figure 6. Ratio ($R$) of the of the shear modulus of NBR/silica nanocomposites [6], characterized by different $\delta_g^I$ values and the shear modulus of net NBR (eq. 12).

It is evident that a more extended super-Arrhenius region (increase in $\delta_g^I$) results in an increase of $R$, i.e., a mechanical reinforcement of the polymer nanocomposites [29,38,40].

## CONCLUSIONS

We have presented in detail a new model for the interpretation of the dynamical and mechanical properties in polymer nanocomposites. Our study is based on a thermodynamic description of the glass transition (two state SLG function [42]). By introducing an Arrhenius component in the glassy region the characteristic $T_g$ is identified. The tests that were performed using the new VFT-like equation in literature data for polymer nanocomposites, indicated that the model can predict $T_g$ values in good agreement with experiment. Moreover, the KK equation provides information regarding the extension of the super-Arrhenius region and the value of the activation energy which governs local dynamics above and below $T_g$.

Based on the temperature dependence of the activation energy, the temperature dependence of the cohesion energy in the glassy region was defined. The SLG EoS which incorporates new mean field ideas, fitted well experimental results of the density in polymer nanocomposites above and below $T_g$ and at different pressures.

The proposed model with relatively simple analytical equations provides new insight for the interpretation of the dynamical and mechanical properties of polymer nanocomposites, as well as for the origin of the glass transition phenomenology in such systems, i.e. the change in the specific volume and the heat capacity step. It highlights the applicability of classical MD simulations, providing tools for a reliable extrapolation in the region of longer relaxation times. By inserting the $T_g$ parameter in the temperature dependence of the effective non-bonded potential, a calibration of CG simulations based on atomistic MD simulations is possible.

SUPPLEMENTARY MATERIAL

Supplementary material (SM) section includes Figure S1, which supports the exponential character of the *g* function, and Figure S2 which fits the temperature dependence of experimental values on PAA providing a validation of the MD simulations analysis on the same system.

AUTHOR INFORMATION

**Corresponding Author**

*E-mail: kritikgio@cheng.auth.gr